# Efficient Electronic Structure Theory via Hierarchical Scale-Adaptive Coupled-Cluster Formalism: I. Theory and Computational Complexity Analysis

Dmitry I. Lyakh*

National Center for Computational Sciences, Oak Ridge National Laboratory, Oak Ridge TN 37831

## Abstract

A novel reduced-scaling, general-order coupled-cluster approach is formulated by exploiting hierarchical representations of many-body tensors, combined with the recently suggested formalism of scale-adaptive tensor algebra. Inspired by the hierarchical techniques from the renormalization group approach, H/H²-matrix algebra and fast multipole method, the computational scaling reduction in our formalism is achieved via coarsening of quantum many-body interactions at larger interaction scales, thus imposing a hierarchical structure on many-body tensors of coupled-cluster theory. In our approach, the interaction scale can be defined on any appropriate Euclidean domain (spatial domain, momentum-space domain, energy domain, etc.). We show that the hierarchically resolved many-body tensors reduce the storage requirements to $O(N)$, where $N$ is the number of simulated quantum particles. Subsequently, we prove that any connected many-body diagram with arbitrary-order tensors, e.g., an arbitrary coupled-cluster diagram, can be evaluated in $O(N\log N)$ floating-point operations. On top of that, we elaborate an additional approximation to further reduce the computational complexity of higher-order coupled-cluster equations, i.e., equations involving higher than double excitations, which otherwise would introduce a large prefactor into formal $O(N\log N)$ scaling.

---

* Corresponding author's emails: liakhdi@ornl.gov, quant4me@gmail.com

This manuscript has been authored by UT-Battelle, LLC under Contract No. DE-AC05-00OR22725 with the U.S. Department of Energy. The United States Government retains and the publisher, by accepting the article for publication, acknowledges that the United States Government retains a non-exclusive, paidup, irrevocable, worldwide license to publish or reproduce the published form of this manuscript, or allow others to do so, for United States Government purposes. The Department of Energy will provide public access to these results of federally sponsored research in accordance with the DOE Public Access Plan (http://energy.gov/downloads/doe-public-access-plan).

# I.     Introduction

The coupled-cluster (CC) theory [1-7] has been one of the most accurate, yet computationally affordable approaches to the electron correlation problem in molecules. Unfortunately, the original formulation cannot be routinely applied to chemical systems with more than $O(100)$ electrons due to a steep polynomial computational scaling pertinent even to the lowest-level approximations, like CCSD [3]. As a consequence, a multitude of approximations to standard coupled-cluster approaches have been elaborated and applied to quite large chemical systems. Basically, these approximations can be roughly grouped into two classes:

- Approximations based on the localization and truncation of many-body interactions, resulting in sparse many-body tensors and (often) linear scaling of the computational cost with respect to the number of correlated particles.

- Approximations based on a low-rank factorization of higher-order many-body tensors by (generally contracted) products of lower-order tensors.

The first class includes the projected-atomic-orbital (PAO) local CC approximation [8,9], the cluster-in-molecule (CIM) local CC approach [10], the fragment molecular orbital (FMO) CC method [11] and closely related method of increments [12], the divide-expand-consolidate (DEC) local CC framework [13], the divide-and-conquer (DAC) CC method [14], the local projected-natural-orbital (LPNO) CC approach [15], the orbital-specific-virtual (OSV) local CC method [16], and some others. The common trait of all these (local) approximations is localization of many-body interactions either in the spatial domain or energy domain, such that the interactions outside of the pre-defined domain are completely neglected, resulting in truncations of many-body tensors. To smooth out discontinuities in calculated molecular properties introduced by such truncations, some methods, like DEC-CC or LPNO-CC, additionally define an error control mechanism (see also Ref. [17]). Despite quite successful application of the truncation-based (local) CC approximations to rather large chemical systems [18], they can still experience an accuracy loss in cases where delocalized quantum effects are significant.

The second class of reduced-scaling CC approximations consists of methods based on the factorization of many-body tensors, including the classical resolution-of-the-identity (ROI) based techniques [19,20], the tensor hypercontraction (THC) CC approach [21], the canonical polyadic (CP) decomposition CC approach [22], the matrix product state (MPS) lower-order CC method [23], and similar formalisms. Normally these approximations do not achieve linear scaling of the computational cost but instead result in some intermediate (reduced) computational complexity, making them less suited for large chemical systems. However, factorization based CC approximations do not require localization of electronic orbitals and truncation of many-body tensors, thus being able to better describe delocalized quantum effects.

On the other side, there has been a tremendous progress in application of the multiresolution analysis (MRA) and related adaptive techniques in electronic structure theory [24-28], although mostly on the self-consistent-field level in practice. In fact, the pragmatic formalism of scale-adaptive tensor algebra [29] used in this paper was inspired by MRA. Very recently, a more elaborated formalism based on the adaptive resolution of many-body tensors was suggested and implemented [30].

Following our long-term goal of constructing an adaptive coupled-cluster theory [31] applicable to arbitrary molecular problems, in this paper we adopt a hierarchical adaptive representation of many-body tensors that allows us to formulate a new reduced-scaling variant of the general-order CC theory (CC theory with arbitrary-order excitations). Essentially, we adopt certain techniques from the renormalization group approach [32-34], multiresolution analysis [24] and H/H$^2$-matrix algebra [35-37], and apply them to general-order many-body tensors within the coupled-cluster formalism. The resulting synthetic general-order CC approach, called HSA@CC (*Hierarchical Scale-Adaptive Coupled-Cluster* formalism), has $O(N)$ memory complexity and $O(N\log N)$ computational complexity, regardless of the highest excitation rank of cluster amplitudes in the CC equations. This is achieved without any truncation of quantum many-body interactions (all parts of the chemical system interact with each other).

In the following sections, we assume that the reader is familiar with the structure of the general-order CC equations [6] while the specific details of the adaptive CC theory and scale-adaptive tensor algebra can be found in Refs. [31] and [29], respectively. In order to proceed to the next section, it is enough to know that the general-order CC equations consist of many-body diagrams [38], each many-body diagram representing either a single many-body tensor or a connected contraction of a finite number of many-body tensors. The latter can always be re-expressed as a sequence of binary tensor contractions. Consequently, in the following discussion we will focus on individual (general-order) coupled-cluster diagrams and binary tensor contractions. Although our problem-specific generalization of certain concepts from hierarchical matrix algebra to tensors will have many overlapping points with the existing general formalism or even with specific applications in quantum many-body theory, like the one from Ref. [28] where a hierarchical multiresolution representation was applied to the coupled-electron pair approximation (CEPA), we still prefer to consistently elaborate our formalism from scratch (as we originally did) for the sake of clarity and due to its specificity to coupled-cluster theory. The elaborated formalism is applicable to any connected coupled-cluster approach, including the Equation-of-Motion (EOM) CC theory [39] and the multireference CC formalism [7] (as long as it is diagrammatically connected).

## II.    Theory

Let us define a general $N$-dimensional *hierarchical vector space* $\boldsymbol{S}$ as an $N$-dimensional vector space with multiple subspaces organized in a tree, the so-called *subspace aggregation tree* (*SAT*). The $N$ leaves of the tree are individual 1-dimensional subspaces the direct sum of which spans $\boldsymbol{S}$ (in other words, the leaves form a basis in $\boldsymbol{S}$). On the next tree level, the leaves are aggregated into larger subspaces represented by the tree nodes on that level. The aggregation is repeated until all subspaces merge into the tree root, the original space $\boldsymbol{S}$, as diagrammatically illustrated in Figure 1 (see the text below for further explanation of the meaning of our diagrams). The 1-dimensional basis subspaces, i.e., the tree leaves, will be enumerated as $\boldsymbol{S}_i, i \in [1:N]$. The aggregate subspaces will be enumerated by specifying an ordered list of the basis subspaces they consist of, or simply the corresponding integer range, for example, $S_{\{1,2,3,4\}}$ or $S_{\{1:4\}}$. Although it is largely irrelevant, in the following discussion the class of vector spaces we deal with will be restricted to the Hilbert spaces as in this work we are interested in solving the molecular Schrödinger equation via the coupled-cluster approach. Thus, the vector space $S$ will have an appropriate inner product and the corresponding induced norm.

Having constructed a specific hierarchical Hilbert space $S$, defined by some subspace aggregation tree, the underlying aggregated subspaces in general may need to be resolved in a lower-quality basis, thus acquiring reduced dimensionality. Following the concepts of *scale-adaptive tensor algebra*\* (SATA) [29], a rectangular transformation matrix $U_{ij}$ (order-2 tensor) will define the mapping from the original subspace to the reduced-dimensional derivative subspace such that $|i| < |j|$, where $|..|$ designates the extent of the corresponding matrix dimension (dimension of the corresponding vector space). In this transformation, the new basis vectors are linearly-independent linear combinations of the original basis vectors. For example, the 4-dimensional aggregate subspace $S_{\{1:4\}}$ can be projected onto some 2-dimensional derivative subspace spanned by two basis vectors defined as linear-combinations of the four original basis vectors via the transformation tensor $U_{ij}$, $|i| = 2, |j| = 4$. Sometimes it is convenient to impose orthogonality on the new (reduced) basis set. In this case, the rows of the rectangular transformation matrix $U_{ij}$ will be orthogonal. Since in this paper we mostly use tensor language, we will say instead that the order-2 transformation tensor $U_{ij}$ is *isometric*, expressed as

$$U_{ij}\overline{U}_{jk} = \delta_{ik}, \tag{2.1}$$

where $\delta_{ik}$ is the Kronecker delta, and $\overline{U}_{jk}$ is a conjugate order-2 tensor defined as $\overline{U}_{jk} = \left(U_{kj}\right)^*$ with * meaning complex conjugation. Since it is common in quantum many-body theory to employ a graphical representation of tensors, Figure 2 illustrates the graphical diagram for the isometric tensor $U_{ij}$. In general, given an arbitrary set (direct sum) of initial vector subspaces and a specific subspace aggregation tree, one can define an isometry at each subspace aggregation point (at each node of the subspace aggregation tree). In such a way, a number of subspaces is aggregated into a larger subspace (direct sum) at each tree node, followed by dimensionality reduction with an appropriate isometric tensor $U_{ij}$. The resulting hierarchical vector space with reduced-dimensional aggregate subspaces has already been represented graphically in Figure 1. The lines at the bottom denote original subspaces the vector space is a direct sum of. Then, the neighboring subspaces are combined and the resulting combined dimension is reduced via an isometric transformation. The entire process is repeated recursively. Note that in this picture any contracted product of isometries is an isometry itself that can easily be checked by contracting any contracted aggregate of isometric tensors with its conjugate and repeatedly applying Eq. 2.1.

---

\* Scale-adaptive tensor algebra is a pragmatic mathematical framework for applications in quantum many-body theory that was originally inspired by the multi-resolution analysis. It is conceptually related to H-matrix algebra, although applied to tensors.

To reiterate, so far we have defined an *N*-dimensional hierarchical Hilbert space that possesses an inner product and the corresponding vector norm. This hierarchical Hilbert space has a tree of subspaces defined on it via the subspace aggregation tree. Each subspace $\boldsymbol{S}_{\{a:b\}}$ has its original dimensionality, $(b - a + 1)$, and, upon necessity, it can also be projected onto some reduced-dimensional subspace $\boldsymbol{S}'_{\{.\}} \subset \boldsymbol{S}_{\{a:b\}}$ by applying an appropriate isometry. Now we will introduce an auxiliary *n*-dimensional Euclidean space $\boldsymbol{D} \equiv \mathbb{R}^n$ with a classical definition of distance. This distance will bring a notion of scale into the hierarchical Hilbert space. Namely, we only need to assume that each original basis vector from $\boldsymbol{S}$ is assigned a bounded-size manifold in $\boldsymbol{D}$ (set of points or analytical shape) centered over a point in $\boldsymbol{D}$. By being of bounded size, we mean that the extent of this manifold over each dimension in $\boldsymbol{D}$ is bounded by some upper limit which can be dimension-specific in general. For the sake of simplicity, we will call this manifold the *auxiliary support* (in $\boldsymbol{D}$) of a basis vector (from $\boldsymbol{S}$). A general requirement on the upper limit of the extent of the auxiliary supports associated with individual basis vectors from $\boldsymbol{S}$ will be that they (support extents) are sufficiently smaller than the combined extent of the auxiliary supports of all basis vectors from $\boldsymbol{S}$. In other words, auxiliary supports in $\boldsymbol{D}$ must be separable to some extent. To make a connection to real physics, one can think of a one-particle Hilbert space in electronic structure theory, that is, a set of one-electron orbitals supported in real 3-dimensional configurational space where the support of each orbital is localized in a specific region of the 3-dimensional space, and these regions combined span over a much larger scale, e.g., over the volume associated with a molecule in this case. In general, it is important to note that the auxiliary Euclidean space does not have to be the real configurational space or even the space over which the specific Hilbert space is defined. It can be any Euclidean (or perhaps even non-Euclidean) space that can be used effectively for constructing sparse hierarchical tensor representations as described below. For example, back to electronic structure theory, the auxiliary space could be the 3-dimensional momentum space if the many-body tensors observe scale separation in that space, or the 1-dimensional energy domain if the many-body tensors observe scale separation in the energy domain (here by scale separation we mean a possibility of partial decoupling of multiple scales when describing particle interactions). The main rationale here is to find a convenient auxiliary space with the following properties:

a) the one-particle basis vectors are associated with generally separable local regions in the chosen auxiliary space;
b) the combined extent of all local regions is significantly larger that the extent of any individual local region;
c) the given many-body operator (many-body tensor) is progressively decaying over the scales of the chosen auxiliary space.

For example, even though the Coulomb operator is decaying in the real configurational space, one will not be able to exploit scale separation if the one-particle vectors are delocalized in the real space as, for example, the canonical Hartree-Fock orbitals are. However, in this case one still may want to exploit separation in the energy domain since each canonical Hartree-Fock orbital is associated with a separable point on the energy scale.

To complete our basic mathematical formalism, we need to back-connect the auxiliary Euclidean space with the hierarchical Hilbert space. Namely, we require the subspace aggregation tree of the hierarchical Hilbert space $S$ to be induced by the configuration of the auxiliary supports in the auxiliary space $D$. We begin with the initial configuration of auxiliary supports in $D$ for all basis vectors from $S$ (each basis vector from $S$ has a local auxiliary support in $D$). As mentioned above, we assume that the auxiliary supports in $D$ are sufficiently separable and are significantly smaller in size than the combined support of them all. Subsequently, the initial auxiliary supports in $D$ can be clustered into larger aggregates recursively. That is, at each step, those auxiliary supports which are close to each other in $D$ will form a cluster. Standard clustering algorithms can be used here. Each cluster in $D$ will define an aggregate subspace in $S$ based on the composition of the cluster and the one-to-one correspondence between basis vectors in $S$ and auxiliary supports in $D$. The respective aggregate subspace in $S$ will be one-to-one associated with the cluster in $D$. By recursively repeating this procedure, all clusters in $D$ will merge and all subspaces in $S$ will be aggregated back into the full Hilbert space $S$. A hierarchical Hilbert space built in this way will be called the *scale-equipped* hierarchical Hilbert space. In general, the clustering procedure will be ambiguous, but our formalism does not require it to not be so. The clustering procedure is only needed to construct a meaningful scale-equipped hierarchical Hilbert space which will form the basis for the hierarchical tensor representation described below. Yet, the clustering procedure is generally expected to keep the clusters compact and maximally separated at each level of clustering.

Having defined a scale-equipped hierarchical $N$-dimensional Hilbert space $S$ (or multiple such spaces), we can construct hierarchical tensor representations by imposing the hierarchical vector space structure on each tensor dimension, somewhat in spirit of the H²-matrix construction [35,36] but for more than two dimensions (generalization to tensors). In the following discussion, a tensor $T_{q_1 \ldots q_m}^{p_1 \ldots p_n}$ maps a value from the Cartesian product of $(n + m)$ integer ranges to a complex number: $T_{q_1 \ldots q_m}^{p_1 \ldots p_n} \to \mathbb{C}$. Here, each tensor index, $p_1 \ldots p_n, q_1 \ldots q_m$, selects a specific basis vector (by its unique integer number) from the vector space associated with the corresponding tensor dimension and the multi-index (tuple) of these indices is mapped to a complex number. Simply restating, the tensor $T_{q_1 \ldots q_m}^{p_1 \ldots p_n}$ is a multi-dimensional array of complex numbers. Should it be necessary, the lower indices represent covariant tensor dimensions and the upper indices represent contravariant tensor dimensions, as in the standard tensor calculus.

Due to the hierarchical structure of vector spaces over which the tensor is defined, and because of the use of adaptive representations [29], each tensor index will acquire a more complex form in our formalism. Specifically, each tensor index will become a triplet $\{s, r, i\} \equiv s^{r:i}$, where $s$ is a subspace from the corresponding hierarchical vector space, $r$ is the specific resolution of that subspace (either full or reduced-dimensional basis living in that subspace), and $i$ is the specific basis vector from subspace $s$ resolved in basis $r$ (either fully or partially). Hence the tensor will acquire the following (a bit hard to read) symbolic representation: $T_{q_1^{r_1:i_1} \dots q_m^{r_m:i_m}}^{p_1^{r_1:i_1} \dots p_n^{r_n:i_n}}$, where the additional subscripts [1..m] and [1..n] enumerate specific tensor dimensions. Fortunately, we will not always need the full symbolic specification, in many cases restricting ourselves back to a simpler form $T_{q_1 \dots q_m}^{p_1 \dots p_n}$, with the difference that now indices $p_1 \dots p_n, q_1 \dots q_m$ refer to the specific subspaces from the hierarchical vector space, not individual basis vectors as before. For the sake of simplicity, we may also abandon the formal distinction between covariant and contravariant tensor indices and only consider the lower indices, e.g., $T_{q_1 \dots q_m}$, unless needed otherwise.

In general, multiple constructive schemes of varying complexity can be considered for building hierarchical tensor representations (akin hierarchical matrix representations). Here we will mostly rely on one of the simplest possible schemes which we call the *recursive diagonal refinement* (RDR scheme). In the RDR scheme, we start from the full tensor $T_{q_1 q_2 \dots q_n}, q_1 = q_2 = \dots = q_n = \boldsymbol{S}$, where each tensor index refers to the full hierarchical vector space $\boldsymbol{S}$ resolved in a reduced basis with resolution $r < r_{max}$. In general, tensor dimensions may refer to different vector spaces but this will not change the formal results below. The full space $\boldsymbol{S}$, which is the root of the subspace aggregation tree, subsequently splits into a direct sum of $m$ smaller subspaces, $\{S_1 \dots S_m\}$, its children in the tree (here $S_k \equiv S_{[a:b]}$, where $[a:b]$ is a unique subrange of the full basis range of $\boldsymbol{S}$). Consequently, the full tensor $T_{q_1 q_2 \dots q_n} \equiv T_{SS \dots S}$ splits into multiple subtensors over the constituent subspaces $\{S_1 \dots S_m\}$: $T_{S_1 S_1 \dots S_1}$, $T_{S_2 S_1 \dots S_1}$, ..., $T_{S_m S_1 \dots S_1}$, $T_{S_1 S_2 \dots S_1}$, $T_{S_2 S_2 \dots S_1}$, ..., $T_{S_m S_2 \dots S_1}$, ..., $T_{S_1 S_m \dots S_1}$, $T_{S_2 S_m \dots S_1}$, ..., $T_{S_m S_m \dots S_1}$, ..., $T_{S_1 S_1 \dots S_m}$, $T_{S_2 S_1 \dots S_m}$, ..., $T_{S_m S_1 \dots S_m}$, $T_{S_1 S_2 \dots S_m}$, $T_{S_2 S_2 \dots S_m}$, ..., $T_{S_m S_2 \dots S_m}$, ..., $T_{S_1 S_m \dots S_m}, T_{S_2 S_m \dots S_m}, ..., T_{S_m S_m \dots S_m}$. Here each children subspace is also generally resolved in a reduced basis with some resolution $r < r_{max}$. The total number of possible subtensors is limited by $m^n$. After the first splitting, subtensors with at least $x$ identical subspaces further split those into their children subspaces (with generally reduced resolution), and the entire procedure is repeated recursively up to the minimally accepted subspace size (ultimately, one may decide to split subspaces all the way up to the individual basis vectors from the original basis). The integer $x$ regulates the principal quality of the hierarchical tensor approximation scheme and the corresponding subscheme is called RDR-$x$. Notice that RDR-1 is equivalent to the regular adaptive dense tiled tensor representation, for example, the one implied in Ref. [29] or the one elaborated in Ref. [30]. Obviously, RDR-1 is not really a hierarchical tensor representation and it does not reduce the asymptotical storage complexity. The truly hierarchical tensor representations start from RDR-2 and their quality decreases with the growing $x$, up to the full tensor order of $n$. Thus, RDR-$n$ is the lowest-quality hierarchical tensor representation scheme for an order-$n$ tensor. In RDR-$n$, only subtensors with all identical indices (fully diagonal subtensors) further split. Note, however, that regardless of the specific hierarchical tensor decomposition scheme, the tensor compression in the RDR-$x$ schemes stems from resolving aggregate subspaces in reduced-dimensional bases. The hierarchical representations only make such a compression scheme scalable with respect to the vector space size.

In the RDR-$x$ schemes with $x>1$, one recursively expands the (many-body) tensor across multiple scales, as defined by the auxiliary space (starting from the coarsest scale and recursively moving to the finer scales). Thus, if the corresponding (many-body) operator is decaying with the increasing scale, one obtains a more balanced and scalable compressed representation of a (many-body) tensor in terms of the Byte/Norm ratio, that is, how much memory is used to store a subtensor with a unity norm. Since the rigorous numerical analysis of the efficiency/error of general hierarchical tensor representations may span multiple specialized papers, here we will restrict ourselves to simple models due to our pragmatic view on how hierarchical tensor representations can be employed in coupled cluster theory in order to construct fast, yet relatively accurate approximations to the electron correlation problem in molecular electronic structure theory. Specifically, for an order-$n$ (many-body) tensor, we will restrict ourselves to the RDR-$x$ schemes, $2 \leq x \leq n$, in which all sufficiently large aggregate subspaces are resolved in reduced bases of size bounded from above by some integer $r_{max}$. By being sufficiently large, we mean that the original subspace dimension is greater than $r_{max}$. Additionally, we will assume that each non-leave node in the subspace aggregation tree always has two or more children, but no more than $b_{max} \geq 2$. In some cases, we will also strengthen this to $b_{max} = 2$ (strictly binary tree) in order to simplify the proofs.

Let us first analyze the asymptotic storage complexity of the RDR-*x* schemes with respect to the dimension $N$ of the one-particle hierarchical Hilbert space $\boldsymbol{S}$ over which the tensor dimensions are defined. For an order-*n* tensor $T_{q_1 \dots q_n}$ (for example, the coupled-cluster amplitude tensor), we begin our analysis with the least expensive RDR-*n* subscheme which we will also call the *fully diagonal refinement* scheme (FDR). As before, the initial tensor is specified over the full Hilbert space $\boldsymbol{S}$ (or multiple such Hilbert spaces in general) and $q_1 = q_2 = \dots = q_n = \boldsymbol{S}$. The full Hilbert space (the root of the subspace aggregation tree) splits into $m \leq b_{max}$ subspaces, resulting in $m^n$ subtensors ($m$ and $n$ do not depend on $N$). Among these, there will be $m$ fully diagonal subtensors ($q_1 = q_2 = \dots = q_n$) which will similarly split further into children subtensors, repeating the entire procedure recursively until reaching the desired subspace granularity level. It is easy to see that all terminal subtensors, that is, the subtensors that do not split further, must have their indices $q_1 \dots q_n$ pointing to the sibling subspaces, that is, the children subspaces of a specific inner node of the subspace aggregation tree. Consequently, the asymptotic number of terminal subtensors will be proportional to the total number of inner nodes in the subspace aggregation tree, which is linear in $N$. Since by construction the full tensor $T_{q_1 \dots q_n}$ is a direct sum of the terminal subtensors and each dimension of each terminal subtensor is resolved in a basis of dimension $r_{max} \sim O(1)$ at most, the total number of Bytes required to store $T_{q_1 \dots q_n}$ is $O(N)$. Consequently, with an appropriate scale-equipped hierarchical Hilbert space, the FDR scheme delivers linear-scaling storage for tensors of decaying many-body operators. The numerical quality of this scheme will depend on the rate of decay of the underlying many-body operator across the scales of the auxiliary space.

Contrary to the FDR scheme, which provides the most memory efficient storage for many-body tensors, other RDR-$x$ schemes with $x < n$ will be less efficient in terms of asymptotic storage complexity for an order-$n$ tensor (but they should be more accurate). Let us try to provide some estimates here. First of all, let us restrict the subspace aggregation tree to a binary tree (for simplicity) and let $(h + 1)$ be its height, that is, the number of tree levels with $h \approx \log_2 N$. Level 0 corresponds to the tree root, level 1 to its children, level 2 to the children of the children, and so on, up to the leave level $h$. In order to estimate the storage complexity for an order-$n$ tensor, we need to estimate the number of terminal subtensors in the RDR-$x$ scheme. Here we will only do this for the RDR-2 scheme which is the most memory consuming one. However, the same formal approach can be used for any other RDR-$x$ scheme. To begin, let us note again that in the RDR-$x$ schemes the full tensor $T_{q_1 \dots q_n}$ is a direct sum of its terminal subtensors (terminal subtensors are pairwise linearly independent by construction). As described before, every terminal subtensor originates from splitting its parental subtensor over two or more identical indices (in RDR-$x$, subtensors with at least $x$ identical indices split further unless the required subspace granularity is reached). We will call the splitting parental subtensor a *prototype subtensor*. The next observation to be made is that each terminal subtensor is obtained only once via a unique path starting from the root prototype tensor and proceeding down through derivative prototype subtensors. We can formalize these paths in the following way. Let us take the tensor multi-index $\{q_1 \dots q_n\}$ and consider index equivalence classes within it. For example, for an order-3 tensor $T_{q_1 q_2 q_3}$ we can distinguish a prototype case where all three indices refer to the same subspace, which we will denote as $\{3\}$. Another prototype case would be $\{2\}\{1\}$ where only two indices refer to the same subspace and the third index refers to another subspace. Finally, $\{1\}\{1\}\{1\}$ would refer to a case when all three indices are pairwise different. Additionally, if two of these three indices are siblings, that is, the corresponding subspaces are children of the same tree node, we will mark this as $\{1\}^2\{1\}$. Or, if all of them are siblings, we will get $\{1\}^3$. Using these definitions, we can now express every possible path how a terminal subtensor could be obtained from the root prototype tensor. For example, let us analyze the storage complexity of the RDR-2 scheme for an order-3 tensor $T_{q_1 q_2 q_3}$. Possible paths are:

1. $\{3\}\rightarrow\{1\}^3$
2. $\{3\}\rightarrow\{2\}\{1\}\rightarrow\{1\}^2\{1\}$

Notice that whenever two or more identical indices of a prototype subtensor split, the corresponding derivative indices will always refer to siblings. That is why $\{1\}\{1\}\{1\}$ is not possible in both paths. Now, in order to count the number of terminal subtensors along each (mutually exclusive) path, we need to count the number of possible prototype subtensors along each path. Let us start from path 1. In this path, a prototype subtensor with all three indices referring to the same (aggregate) subspace splits into terminal subtensors each of which does not have a single pair of identical indices. This splitting can occur at (almost) any inner node of the subspace aggregation tree. Thus, the number of prototype subtensors along this path will be proportional to the number of inner tree nodes, $O(N)$. The number of terminal subtensors is $O(1)$ related to the number of their respective prototype subtensors as each prototype subtensor always splits into $O(1)$ derivative subtensors. Consequently, the total number of terminal subtensors and storage complexity in path 1 is $O(N)$.

Path 2 is a bit more complicated as there are two different prototype subtensors involved, $\{3\}$ and $\{2\}\{1\}$. First of all, similarly to path 1, the prototype subtensor of form $\{3\}$ can refer to (almost) any inner node of the subspace aggregation tree. Specifically, at tree level $L<h$, there are $2^L$ inner nodes, and, consequently, the same number of prototype subtensors of form $\{3\}$ are possible. Then, the split $\{3\}\rightarrow\{2\}\{1\}$ can occur at any descendant inner node (level > $L$) the total number of which is bounded by $O(2^{(h-L-1)})$. The number of possible combinations is $O\left(2^L \cdot 2^{(h-L-1)}\right) = O(2^h)$. Summing over all tree levels, $0 \le L < h$, we will get

$$\sum_{L=0}^{h-1} O(2^h) = O(h \cdot 2^h)$$

Since $h \approx \log_2 N$, the final storage complexity for path 2 and for the full RDR-2 scheme for an order-3 tensor is $O(N\log_2 N)$ Bytes.

Analogously, we can determine the asymptotic storage complexity for higher-order tensors and all applicable RDR-*x* schemes. Let us give just one more example for an order-4 tensor and RDR-2. There are more paths possible for subtensor splitting in this case:

1. $\{4\}\rightarrow\{1\}^4$
2. $\{4\}\rightarrow\{3\}\{1\}\rightarrow\{1\}^3\{1\}$
3. $\{4\}\rightarrow\{3\}\{1\}\rightarrow\{2\}\{1\}\{1\}\rightarrow\{1\}^2\{1\}\{1\}$
4. $\{4\}\rightarrow\{2\}^2\rightarrow\{1\}^2\{1\}^2$
5. $\{4\}\rightarrow\{2\}^2\rightarrow\{2\}\{1\}^2\rightarrow\{1\}^2\{1\}^2$
6. $\{4\}\rightarrow\{2\}\{1\}^2\rightarrow\{1\}^2\{1\}^2$

Let us analyze path 4. Similarly to previous cases, the prototype subtensor $\{4\}$ can refer to (almost) any inner tree node. Let us consider a specific tree level $L<h$ with $2^L$ candidate inner nodes. The subsequent split $\{2\}^2\rightarrow\{1\}^2\{1\}^2$ (actually two independent splits $\{2\}\rightarrow\{1\}^2$) can occur at any pair of descendant inner nodes with $O(2^{(h-L-1)})$ candidates each and $O(2^{2(h-L-1)})$ combinations. Summing over all tree levels, we get

$$\sum_{L=0}^{h-1} 2^L O\big(2^{2(h-L-1)}\big) = \sum_{L=0}^{h-1} O\big(2^L \cdot 2^{2(h-L)}\big) = \sum_{L=0}^{h-1} O(2^{2h-L}) = O(h \cdot 2^{2h})$$

Since $h \approx \log_2 N$, we end up with $O(N^2 \log_2 N)$ Bytes for path 4, which is super-quadratic in $N$. Thus, the RDR-*x* schemes other than FDR do not generally guarantee linear scaling of tensor storage. Consequently, the FDR scheme will be central in constructing fast coupled-cluster theory as shown below. To analyze other RDR-*x* schemes, one will need to keep only those paths which can be realized within the chosen RDR-*x* scheme.

As known, coupled-cluster theory [6,7] provides a general framework for building practical size-extensive [40] approximations to the exact many-body wavefunction. Each such an approximation is represented by a set of tensor equations and a specific solution procedure for finding the unknown cluster amplitude tensors (and other relevant tensors). Traditionally, the tensor equations of coupled-cluster theory are decomposed into the so-called many-body diagrams, each many-body diagram being either a tensor or a tensor contraction of two or more tensors. Additionally, a properly formulated coupled-cluster approach should solely consist of connected many-body diagrams only, meaning that the underlying tensor contractions cannot contain a subgroup of tensors fully contracted to a scalar [41]. Based upon a regular (flat) one-particle Hilbert space, the computational complexity of solving the coupled-cluster equations quickly hits a steep polynomial wall, thus restricting the class of feasible many-body systems to $O(100)$ particles at best (on current HPC platforms). By replacing the flat dense tensor representation with the hierarchical FDR scheme elaborated above, we show below that any connected coupled-cluster formalism will require only $O(N\log N)$ Flops regardless of the highest tensor order present in the tensor equations, thus delivering a new fast version of the general-order coupled-cluster theory without interaction cut-offs.

**Proposition 1**. Having imposed the hierarchical FDR representation on many-body tensors, any connected coupled-cluster diagram can be evaluated in $O(N\log N)$ Flops, where $N$ is the size of the one-particle basis.

**Proof**

A connected coupled-cluster diagram generally requires evaluation of one or more binary tensor contractions, each with at least one contracted dimension (labelled by a repeated index). If we assume a finite number of binary tensor contractions, it is sufficient to show that any binary tensor contraction operating on hierarchical tensors constructed with the FDR scheme requires $O(N\log N)$ Flops. A general form of a binary tensor contraction is (implicit summation over repeated tensor indices is implied)

$$D_{\{l_1 \ldots l_m\},\{r_1 \ldots r_n\}} = L_{\{l_1 \ldots l_m\},\{c_1 \ldots c_k\}} R_{\{c_1 \ldots c_k\},\{r_1 \ldots r_n\}}$$

where the tensor indices are logically grouped into three groups, a group of contracted indices, $\{c_1 \ldots c_k\}$, a group of uncontracted indices from the left tensor argument, $\{l_1 \ldots l_m\}$, and a group of uncontracted indices from the right tensor argument, $\{r_1 \ldots r_n\}$ (so far we assume a flat representation of the underlying vector spaces with each index enumerating individual basis vectors). Any binary tensor contraction can be represented in this form by means of index permutations. In our formalism, all participating tensors are constructed hierarchically with the FDR scheme such that each tensor is essentially a direct sum of its terminal subtensors. By construction, the extent of each dimension of each terminal subtensor is less or equal to some constant integer $r_{max}$. By switching from a flat vector space representation to the hierarchical one, we have to relax the above general tensor contraction form to

$$\widetilde{D}_{\{l_1 \dots l_m\},\{r_1 \dots r_n\}} = \widetilde{L}_{\{l'_1 \dots l'_m\},\{c_1 \dots c_k\}} \widetilde{R}_{\{c'_1 \dots c'_k\},\{r'_1 \dots r'_n\}}$$

where now all indices designate subspaces from the hierarchical vector space. The primed indices designate subspaces *related* to the corresponding unprimed indices, that is, each primed index may refer to the same subspace as its unprimed counterpart, or any of its ancestors, or any of its descendants. In other words, primed and unprimed subspaces must have a non-zero overlap (in fact, this condition becomes the principal condition in non-orthogonal bases where overlapping subspaces do not have to be genealogically related). The relaxed general tensor contraction form can easily be obtained from the original tensor contraction form by mapping subtensors from the hierarchical FDR representation back to the flat representation:

$$U_{l_1}^{l'_1} \dots U_{l_m}^{l'_m} \widetilde{D}_{\{l'_1 \dots l'_m\},\{r'_1 \dots r'_n\}} U_{r_1}^{r'_1} \dots U_{r_n}^{r'_n} = U_{l_1}^{l''_1} \dots U_{l_m}^{l''_m} \widetilde{L}_{\{l''_1 \dots l''_m\},\{c'_1 \dots c'_k\}} \times$$

$$\times U_{c_1}^{c'_1} \dots U_{c_k}^{c'_k} \overline{U}_{c_1}^{c''_1} \dots \overline{U}_{c_k}^{c''_k} \widetilde{R}_{\{c'_1 \dots c''_k\},\{r'_1 \dots r''_n\}} U_{r_1}^{r'_1} \dots U_{r_n}^{r''_n}$$

where the isometric tensors $U$ expand reduced-dimensional aggregate subspaces back to the full flat space (by contracting all repeated indices, one will obtain back the original general tensor contraction form). Now, by contracting both sides, pre-multiplied with the conjugated isometric tensors $\overline{U}_{l_1}^{l'_1} \dots \overline{U}_{l_m}^{l'_m} \overline{U}_{r_1}^{r'_1} \dots \overline{U}_{r_n}^{r'_n}$, over all unprimed indices, we obtain

$$\widetilde{D}_{\{l'_1 \dots l'_m\},\{r'_1 \dots r'_n\}} = S_{l'_1}^{l''_1} \dots S_{l'_m}^{l''_m} \widetilde{L}_{\{l''_1 \dots l''_m\},\{c'_1 \dots c'_k\}} S_{c'_1}^{c''_1} \dots S_{c'_k}^{c''_k} \widetilde{R}_{\{c'_1 \dots c''_k\},\{r'_1 \dots r''_n\}} S_{r'_1}^{r''_1} \dots S_{r'_n}^{r''_n}$$

where the order-2 tensors $S$ are the overlap matrices between subspaces of the hierarchical vector space, and the tensors $\widetilde{D}$, $\widetilde{L}$, and $\widetilde{R}$ are hierarchical tensors composed of subtensors constructed with the FDR scheme. Consequently, tensor $\widetilde{L}$ consists of $O(N)$ subtensors. In each of these subtensors, the multi-index $\{c'_1 \dots c'_k\}$ always refers to the sibling subspaces, that is, the children of a specific aggregate subspace $S_p$ (in FDR, all indices of any subtensor must refer to the sibling subspaces, unless they all refer to the root space). Let us assume for now that all basis subspaces are pairwise orthogonal and the subspace aggregation tree is binary. Then, for a given $S_p$, which refers to a specific node of the subspace aggregation tree, the multi-index $\{c''_1 \dots c''_k\}$ from the $\widetilde{R}$ tensor must either have all its indices refer to an ancestor node of $S_p$ (or $S_p$ itself) or all of them be children of a descendant node of $S_p$ (also including $S_p$). Any other valid value of $\{c_1'' \ldots c_k''\}$ will necessarily result in a zero overlap in the FDR scheme: $S_{c_1}^{c_1''} \ldots S_{c_k}^{c_k''} = 0$. In the first case, for any $S_p$, the number of the ancestor nodes (subspaces) is $O(\log N)$ bounded because of the tree structure, resulting in $O(N \log N)$ total combinations of the $\tilde{L}$ and $\tilde{R}$ subtensors. For the second case, let us consider a subspace $S_p$ at tree level $L < h$, in which case the number of descendant nodes is $O(2^{h-L})$ bounded. Summing over all tree levels ($2^L$ nodes at tree level $L$ with $O(2^{h-L})$ descendants each), we will get the total number of possible combinations of the $\tilde{L}$ and $\tilde{R}$ subtensors to be

$$\sum_{L=0}^{h-1} 2^L O(2^{h-L}) = \sum_{L=0}^{h-1} O(2^L 2^{h-L}) = \sum_{L=0}^{h-1} O(2^h) = h \cdot O(2^h) = O(h \cdot 2^h)$$

where $h \approx \log_2 N$, resulting in $O(N \log N)$.

In both cases above, any contracted combination of the $\tilde{L}$ and $\tilde{R}$ subtensors contributes to a specific $\tilde{D}$ subtensor by projecting uncontracted multi-indices $\{l_1'' \ldots l_m''\}$ and $\{r_1'' \ldots r_n''\}$ onto the $\{l_1' \ldots l_m'\}$ and $\{r_1' \ldots r_n'\}$ subspaces, respectively. The projection requires $O(1)$ Flops as the dimensions of the overlap matrices are bounded in size by a constant integer $r_{max}$ (maximal reduced dimension of any subspace). The extents of dimensions of all subtensors are bounded by the same integer for the same reason. Consequently, there are $O(N \log N)$ valid combinations of the $\tilde{L}$ and $\tilde{R}$ subtensors each of which can be evaluated in $O(1)$ Flops, resulting in $O(N \log N)$ Flops in total.

In the above proof, we made two intermediate assumptions: (a) pairwise orthogonality of the basis subspaces, and (b) a binary subspace aggregation tree. Let us now extend the proof beyond these assumptions. Specifically, we will assume that the basis subspaces are not necessarily pairwise orthogonal, however, each subspace may have a non-zero overlap only with a bounded number of other subspaces on the same level of the subspace aggregation tree which is no longer binary (but the tree node branching factor is still bounded). Again, any $\tilde{L}$ subtensor has its multi-index $\{c_1' \ldots c_k'\}$ refer to sibling subspaces such that this multi-index can be uniquely characterized by the parental subspace $S_p$ of the sibling subspaces included in $\{c_1' \dots c_k'\}$, unless all indices in $\{c_1' \dots c_k'\}$ refer to the root space. Then, given a specific $\tilde{L}$ subtensor, the parental subspace of the sibling multi-index $\{c_1'' \dots c_k''\}$ from the $\tilde{R}$ subtensor must have a non-zero overlap with $S_p$, otherwise the overlap matrices will be zero. Consequently, for each $S_p$, we need to count the number of the inner tree nodes (plus the root node) for which the overlap with $S_p$ is non-zero. Similarly to the orthogonal case considered above, all ancestor nodes of $S_p$ (including $S_p$ itself) qualify. The number of the ancestor nodes is bounded by the tree height $h \leq (\log_2 N + 1)$, thus still giving the same $O(N\log N)$ Flop scaling. Besides the ancestor nodes, the set of descendant nodes of $S_p$ will contain subspaces that overlap with $S_p$. Additionally, since $S_p$ may now overlap with $O(1)$ nodes at the same tree level, the set of all descendant nodes of those will also contain subspaces that overlap with $S_p$. Consequently, the total number of combinations of the $\tilde{L}$ and $\tilde{R}$ subtensors for which the overlap matrices are not identically zero is bounded from above by

$$\sum_{S_p} \sum_{S_q \cap S_p \neq 0} 1$$

where the first summation runs over all inner tree nodes (subspace $S_p$), and the second summation runs over all subspaces ($S_q$) which are located at the same or lower tree level and which may have a non-zero overlap with $S_p$. In this double summation, each subspace $S_q$ may have a non-zero overlap with at most $h(1 + t)$ subspaces, where $h \leq (log_2 N + 1)$ is the tree height and $t$ is the upper bound to the number of the lateral subspaces the subspaces $S_q$ is allowed to overlap with on the same tree level. For each of those subspaces, $S_q$ will also overlap with all their ancestors whose number is less or equal to $ht$. Since $t \sim O(1)$ by our assumption and the range of the first summation is $O(N)$, we still end up with the same $O(N\log N)$ Flop scaling.

**End of proof**

Despite the fact that any connected coupled-cluster diagram with hierarchical tensors constructed with the FDR scheme can be asymptotically evaluated in $O(N\log N)$ Flops, coupled-cluster diagrams involving higher-order tensors will necessarily introduce large prefactors in this formal scaling. In order to reduce these prefactors, we can introduce another approximation that is also based on the isometric compression of the underlying vector spaces. To this point, our isometric transformations projected a given vector space into one of its subspaces, thus reducing the dimensionality. We call such an isometry the 1-1 isometry, represented by an order-2 isometric tensor (1-1 isometric tensor). For higher-order tensors, we can introduce a generalization, namely, the $n$-1 isometry which projects a direct product of $n$ vector spaces, which is a vector space itself, into one of its subspaces. For example, if $\boldsymbol{S}^{(1)}, \dots, \boldsymbol{S}^{(n)}$ are vector spaces, we can introduce an isometric map

$$\boldsymbol{S}^{(1)} \otimes \dots \otimes \boldsymbol{S}^{(n)} \mapsto \boldsymbol{V}, \qquad \boldsymbol{V} \subset \boldsymbol{S}^{(1)} \otimes \dots \otimes \boldsymbol{S}^{(n)}$$

represented by an order-$(n+1)$ tensor $U_{q,p_1 \dots p_n}$ in which the multi-index $\{p_1 \dots p_n\}$ enumerates basis vectors of $(\boldsymbol{S}^{(1)} \otimes \dots \otimes \boldsymbol{S}^{(n)})$ and index $q$ refers to the basis vectors of $\boldsymbol{V}$ ($\boldsymbol{V}$ does not have to possess a direct-product structure in general). Such an isometric transformation is quite common in the renormalization group techniques [42], in particular, in the tensor network state (TNS) theory [33,34]. Not only it reduces the dimensionality of the space, but also the tensor order as $n$ tensor indices are replaced by a single one. Below we show that this transformation is also useful for reducing the complexity of higher-order coupled-cluster equations. A somewhat similar in spirit technique was used previously in Refs. [43,44] in order to reduce the computational scaling of the most expensive terms in the CCSD and CCSDT equations.

Since in essence we adapt the isometric compression technique from tensor network state theory [33,34] to the general-order coupled-cluster approach, below we will use a graphical (diagrammatic) representation of many-body tensors, which is common in tensor network state theory, with specific modifications dictated by the coupled-cluster ansatz. Similarly to others, in our representation an order-$n$ tensor is depicted by a vertex (or a shape) with $n$ legs, each leg corresponding to a specific tensor dimension. Tensor operations are formed by combining graphical tensors together and possibly contracting some (or all) their legs. Namely, if specific legs of two or more tensors are combined, a summation over the corresponding dimensions of the participating tensors is implied. Figure 3 illustrates some basic examples.

In coupled-cluster theory, it is common to have a specific vacuum state, thus distinguishing between hole and particle states (hole-particle second quantization) [6]. As a consequence, coupled-cluster tensors may have two kinds of legs, those corresponding to the holes and those corresponding to the particles, as captured by the standard coupled-cluster diagram techniques. In our discussion, we will use a synthetic graphical representation (diagram technique) derived from the Hugenholtz representation. Specifically, the many-body tensors forming a particular coupled-cluster diagram are positioned in-order along a horizontal line, the vacuum line. All hole legs (lines) are below the vacuum line while all particles legs (lines) are above the vacuum line; legs corresponding to second-quantized creators are directed to the left while legs corresponding to second-quantized annihilators are directed to the right, thus the latter being able to combine (contract) with the former from left to right, reflecting the standard Wick's contraction rule [6]. An example of a coupled-cluster diagram expressed in our graphical representation in shown in Figure 4. The corresponding equation is

$$P(a_1|a_2a_3)P(i_1i_2|i_3)H_{c_1c_2}^{k_1k_2}T_{k_1i_1i_2}^{c_1c_2a_1}T_{k_2i_3}^{a_2a_3}$$

where $H_{c_1c_2}^{k_1k_2}$ is the two-body Hamiltonian tensor, $T_{k_2i_3}^{a_2a_3}$ and $T_{k_1i_1i_2}^{c_1c_2a_1}$ are cluster amplitudes, while $P(a_1|a_2a_3)$ and $P(i_1i_2|i_3)$ are index permutation generators (anti-symmetrizers).

In the flat representation of the underlying vector spaces, each tensor index (tensor leg) enumerates individual basis vectors from the corresponding vector space. If the dimension of each space is $O(N)$ bounded, the asymptotical cost of the numerical evaluation of a binary tensor contraction will be $O(N^{k+l})$, where $k$ is the number of uncontracted indices (legs) and $l$ is the number of contracted indices (legs) without repeats. By switching to the hierarchical tensor representation constructed with the FDR scheme, the asymptotic computational cost of diagram evaluation is reduced to $O(N\log N)$, presenting already enormous savings for connected diagrams with higher-order tensors. Yet, the diagram evaluation prefactor may still grow with the order of participating tensors. Consequently, one can benefit from an approximation in which the order of the tensors is reduced. This is exactly where we can benefit from an $n$-1 isometric transformation. Specifically, each higher-order tensor can be contracted with one or more $n$-1 isometric tensors, thus shrinking its order to the desired level. Note that in the hierarchical tensor representation each tenor index is resolved in multiple stages, subspace → reduced basis → basis vector, such that a summation over an index splits into two nested summations, one over the subspaces and the other one over the basis vectors of a specific reduced basis of a specific subspace. Figure 5 illustrates a number of $n$-1 isometries and multiple examples of tensor order reduction for order-4 and higher-order tensors. The quality of such a compression can be assessed by back projecting the compressed dimensions into the original (direct-product) space. For an $n$-1 isometry $U_{q,p_1\ldots p_n}$, the corresponding compressing projector (*compressor*) is

$$P_{q_1\ldots q_n}^{p_1\ldots p_n} \equiv \overline{U}^{p_1\ldots p_n, r} U_{r, q_1\ldots q_n}$$

By applying compressors to all relevant tensor dimensions with a subsequent evaluation of the norm of the compressed tensor and its difference from the norm of the original tensor, one can assess the error introduced by the compression. Obviously, the optimal isometric tensors will be specific to the tensor for which the compression is done. For order-2 tensors, which can be mapped to matrices uniquely, the optimal compression is provided by the singular value decomposition (in the Frobenius norm). For higher-order tensors some problem-specific heuristics should be used in general.

In general, given an arbitrary connected many-body (e.g., coupled-cluster) diagram, we proceed as follows (see Figure 6 for a graphical illustration). First, we apply a chosen set of $n$-1 isometries to the residual tensor, that is, the tensor obtained by performing all index contractions (Figure 6a). Second, we apply a chosen set of compressors to each participating input tensor (Figure 6b). In both cases, we first group tensor dimensions into groups, subsequently applying an isometry (for the residual tensor) or a compressor (for input tensors) to each group of dimensions we want to compress. Then, each input tensor absorbs the closest isometric tensor from each compressor applied to it by contracting with it. At this point, the residual (output) tensor and all input tensors have been converted into the compressed representation. However, we still may have remaining isometric tensors in our many-body diagram (Figure 6c). To finalize our procedure, we group the remaining isometric tensors into a maximally possible number of groups such that the contraction of the isometric tensors within each group completely gets rid of all diagram lines attached to the bottom of any isometric tensor (bottom of the corresponding triangle). The contracted products of isometric tensors obtained in this way will be called *connectors* (Figure 6d). The final (converted) many-body diagram will consist of compressed tensors only and, possibly, connectors. As the number of diagram lines is reduced, the computational scaling prefactor will be reduced as well, for the price of a more compressed (approximate) representation of the underlying many-body tensors. Consequently, in our formalism we introduce two major approximations:

a) Hierarchical representation of many-body tensors constructed with the FDR scheme.

b) Additional compression of multiple tensor dimensions into a single (effective) tensor dimension.

Importantly, both approximations are adaptive, that is, their quality can be adjusted as dictated by the problem in hand. This can always be done by varying the dimensionality of all involved isometric tensors in both approximations. In the full dimension limit, all isometric tensors will become unitary, thus introducing no associated numerical error, but also not providing any Flop or memory savings. The analysis of the practical performance and efficiency of our formalism is deferred to future works as the corresponding computer implementation would require a rather non-trivial effort. At this point, we will only provide a general algorithm how to apply our formalism to any given (connected) coupled-cluster approximation in order to reduce its computational scaling to an $O(N\log N)$ asymptotic cost:

1. Given a finite one-particle Hilbert space(s), introduce physically motivated auxiliary Euclidean space(s) which will guide the subspace aggregation in the Hilbert space(s) into the subspace aggregation tree(s) such that progressively more sparse tensor representations can be deployed at increasingly large scales.

2. Define an initial set of reduced bases for all subspaces of the hierarchical Hilbert spaces(s) by recursively combining multiple subspaces according to the subspace aggregation tree(s) and subsequently reducing the dimension of the combined space via a 1-1 isometric transformation. The corresponding 1-1 isometric tensors are specifically chosen/optimized to minimize the norm loss for the relevant (prechosen) tensor(s). At this point, all many-body tensors are expressed in a hierarchical representation (e.g., FDR).

3. If higher-order tensors are present in the many-body diagrams of the method of interest, convert the relevant many-body diagrams into the reduced-order representation by introducing $n$-1 isometric transformations. The corresponding $n$-1 isometric tensors are specifically chosen/optimized to minimize the norm loss for the relevant (prechosen) tensor(s).

4. Solve the final set of (compressed) equations based on the compressed hierarchical tensors.

5. Optionally, adjust all relevant isometric tensors to decrease the approximation error and go back to step 2. Repeat until satisfaction.

## Conclusions

We have presented a rather general technique that can be used for reducing the memory and computational complexity of an arbitrary (connected) coupled-cluster approximation to $O(N)$ and $O(N\log N)$, respectively. The technique is equally applicable to the EOM-CC extension for excited electronic states and multireference CC theory. Most notably, the presence of higher-than-double (actually, arbitrary-order) excitations in the CC equations does not change the formal asymptotic scaling, although it may introduce a large prefactor. Our technique is inherently adaptive as higher-order cluster excitations together with new cluster excitations of the same order can be included gradually, as needed by the described electronic state (while monitoring the convergence of the calculated property). Currently, we are actively working on the implementation of our technique in order to benchmark its efficiency in future.

## Acknowledgments

This research used resources of the Oak Ridge Leadership Computing Facility, which is a DOE Office of Science User Facility supported under Contract DE-AC05-00OR22725.

**Figure 1**

Diagrammatic representation of a hierarchical vector space. Each connecting line is associated with a specific subspace from the hierarchical vector space. The direct-sum sign combines two subspaces together. The isometric transformation represented by a triangle with one incoming and one outgoing line projects the combined subspace into a reduced-dimensional subspace.

Diagrammatic space coarsening:

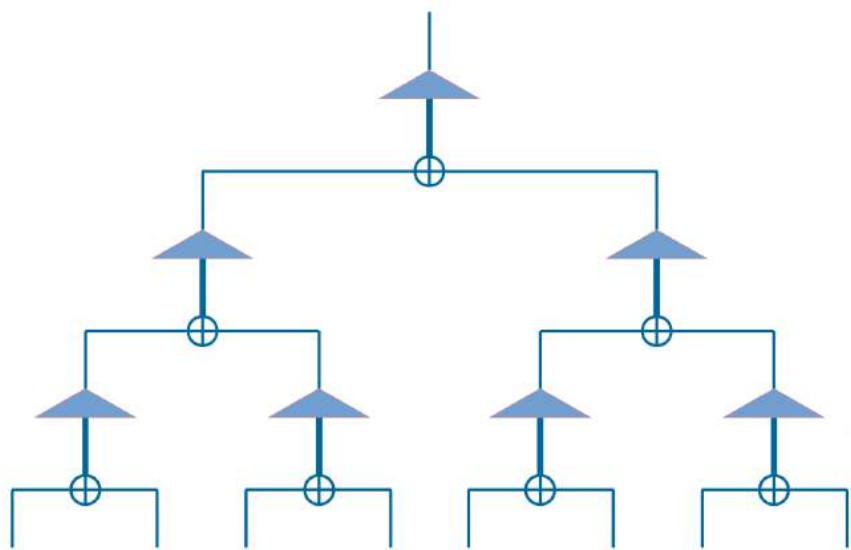

**Figure 2**

Isometric tensor and isometric condition. The thicker line represents a higher-dimensional subspace than the thinner line. The free line on the right is the Kronecker delta in the reduced-dimensional subspace.

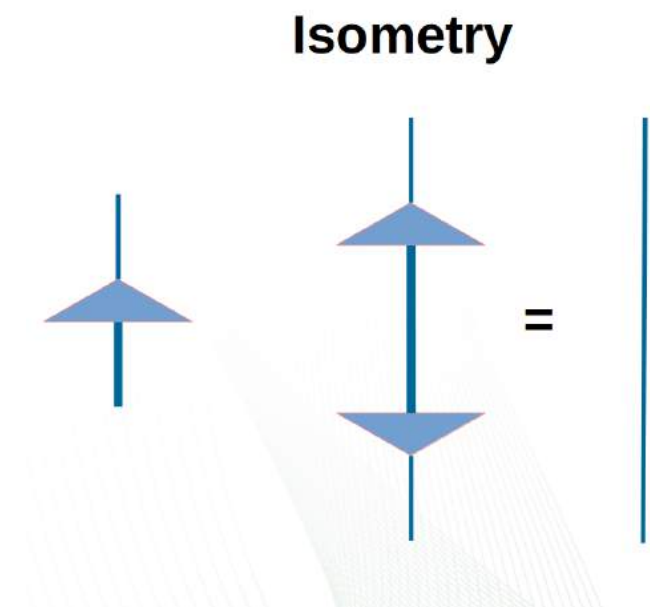

**Figure 3**

Diagrammatic representation of basic linear algebra objects and operations.

Diagrammatic tensor representation:

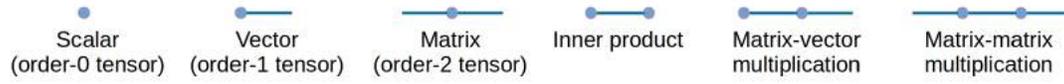

| Scalar | Vector | Matrix | Inner product | Matrix-vector | Matrix-matrix |
|---|---|---|---|---|---|
| (order-0 tensor) | (order-1 tensor) | (order-2 tensor) | | multiplication | multiplication |

**Figure 4**

An example of a coupled-cluster diagram in our graphical representation.

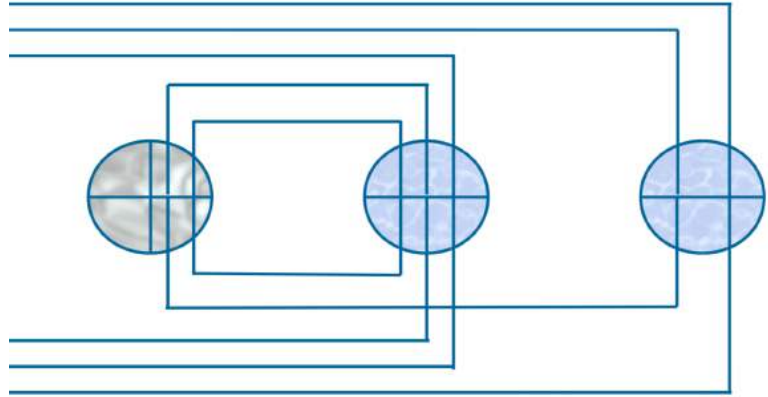

**Figure 5**

Higher-order isometric transformations and isometric conditions used to reduce the order of many-body tensors. Note the necessity of the antisymmetrization in general.

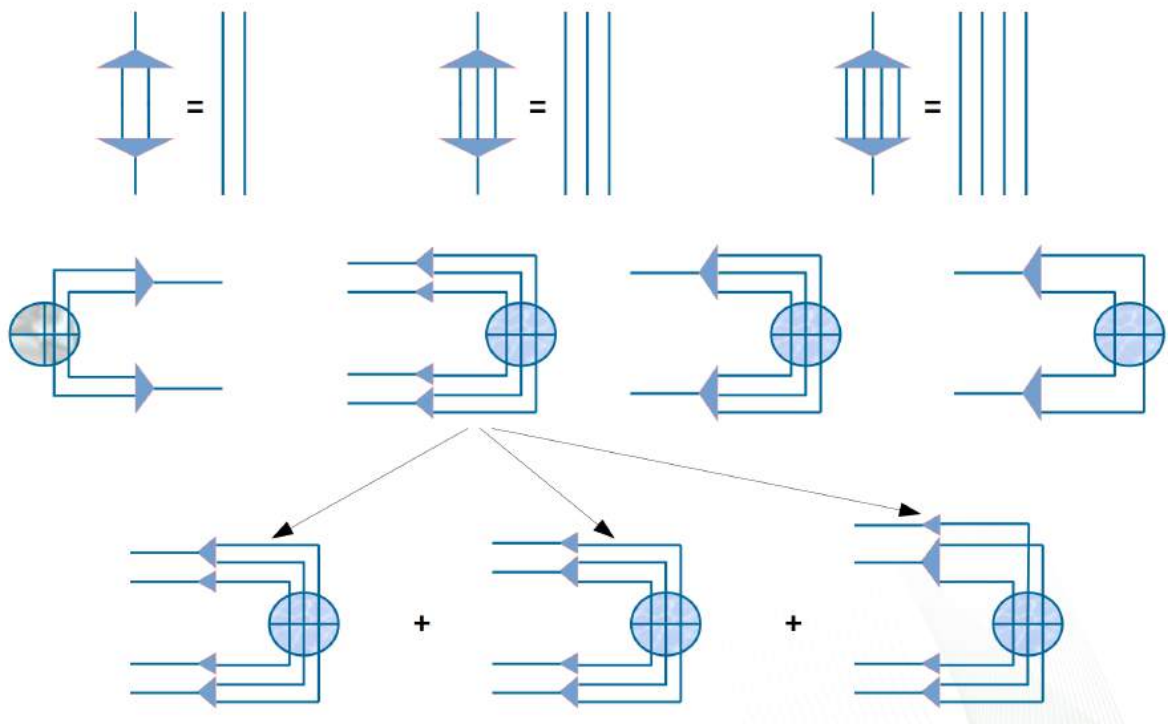

**Figure 6**

Tensor order reduction (compression) applied to a coupled-cluster diagram (see text for details).

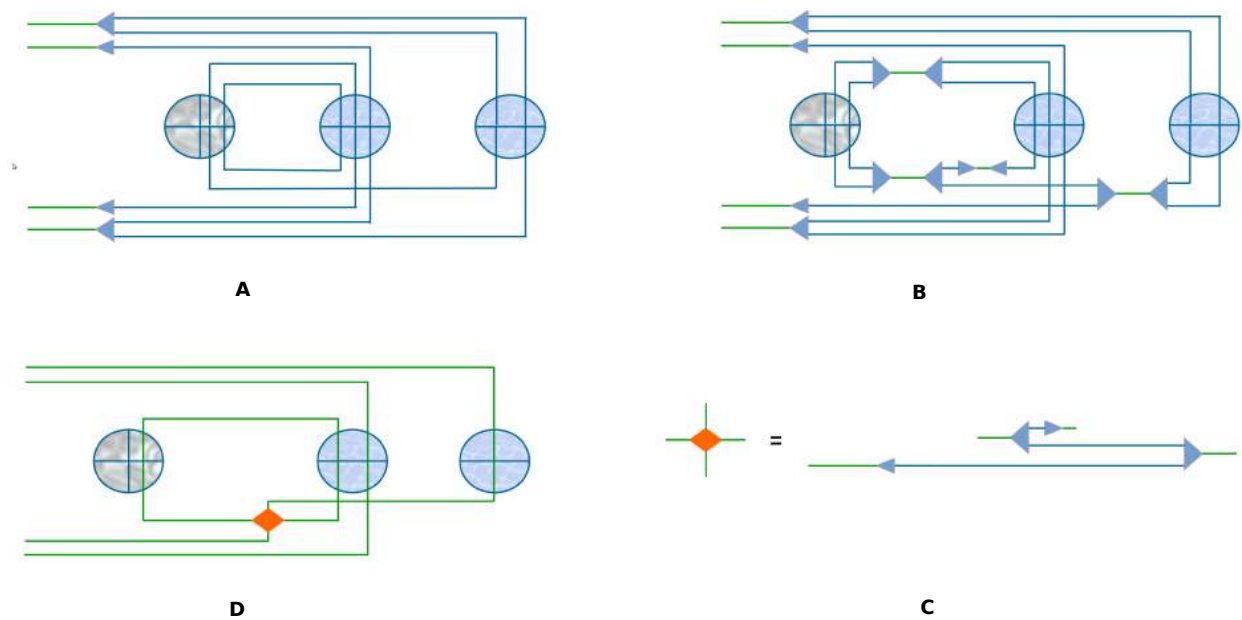